\documentclass[reqno,11pt]{amsproc}     

\usepackage[utf8]{inputenc}
\usepackage[round]{natbib} 
\usepackage{graphicx,float,epstopdf,multirow}
\usepackage{amsmath,amsfonts,amssymb,amsaddr,enumitem}
\usepackage{adjustbox,tabularx,array,booktabs,ragged2e,siunitx}
\usepackage[a4paper, margin=0.8in,headsep=.4in]{geometry}

\usepackage[table]{xcolor}
\definecolor{lightcyan}{rgb}{0.88, 1.0, 1.0}
\definecolor{lemonchiffon}{rgb}{1.0, 0.98, 0.81}
\definecolor{bananamania}{rgb}{0.98, 0.91, 0.71}
\definecolor{lightgreen}{rgb}{0.56, 0.93, 0.56}

\usepackage{pgfplots,pdfpages}
\usepackage{caption}
\usepackage{subcaption}

\usepackage{doc}
\makeatletter
\def\@setthanks{\vspace{-\baselineskip}\def\thanks##1{\@par##1\@addpunct.}\thankses}
\makeatother

\title{Navigating simplicity and complexity of social-ecological
systems through a dialogue between dynamical systems
and agent-based models}

\author[Sonja]{Sonja Radosavljevic$^{1,*}$, Udita Sanga$^{1,2}$, Maja Schlüter$^1$}\thanks{$^*$ Corresponding author: Sonja Radosavljevic, sonja.radosavljevic@su.se  \\\hspace*{0.8em} $^1$ Stockholm Resilience Centre, Stockholm University \\\hspace*{0.8em} $^2$ Friedman School of Nutrition Science and Policy, Tufts University}

\begin{document}

\maketitle

\begin{abstract}
Social-ecological systems research aims to understand the nature of social-ecological phenomena, to find ways to foster or manage conditions under which desired phenomena occur or to reduce the negative consequences of undesirable phenomena. Such challenges are often addressed using dynamical systems models (DSM) or agent-based models (ABM). Here we develop an iterative procedure for combining DSM and ABM to leverage their strengths and gain insights that surpass insights obtained by each approach separately. The procedure uses results of an ABM as inputs for a DSM development. In the following steps, results of the DSM analyses guide future analysis of the ABM and vice versa. This dialogue, more than having a tight connection between the models, enables pushing the research frontier, expanding the set of research questions and insights. We illustrate our method with the example of poverty traps and innovation in agricultural systems, but our conclusions are general and can be applied to other DSM-ABM combinations.
\end{abstract}

%% \linenumbers

%% main text
\section{Introduction}

One of the main aims of social-ecological systems (SES) research is to understand the nature of social-ecological phenomena. This knowledge is important for finding ways to foster conditions under which desirable phenomena occur, but also for finding ways to change conditions that lead to undesirable phenomena and reduce their negative consequences. Research often relies on modeling to capture social-ecological interactions and explore the dynamics that arise from them \citep{Schluter2019b}. \citet{Anderies} emphasize that SES models need to include ecological and social components without downplaying the importance of either. Reasons for this can be found in characteristics of people and human societies that make them different from other organisms and ecological communities. For example, the capacity of people to act intentionally, to use technology or act on wider spatial, temporal and organizational levels in comparison to other organisms cannot be fully represented using classical ecological models. Similarly, models of human behavior in economic or bioeconomic models used in resource management often ignore ecological complexity and reduce it to a single variable \citep{Epstein}. Some neglected ecological dynamics can be critical for gaining a deep understanding of SES.

Dynamical system models (DSM) and agent based models (ABM) are frequently used in SES research.. DSM are designed to study evolution of a system from its initial state according to the rule of evolution \citep{Radosavljevic2023}. They are usually defined as systems of ordinary or partial differential equations and the analysis is focused on uncovering attractors and conditions for their existence. This is closely related to exploring asymptotic dynamics, stability and bifurcations, making DSM a useful tool in studying regime shifts \citep{Lade2013, Martin}, effects of social norms on renewable resource management \citep{Tavoni}, poverty traps \citep{Lade2017, Ngonghala, Radosavljevic2020, Radosavljevic2021}, self-organized governance in coupled infrastructure systems \citep{Muneepeerakul} or social-ecological transitions \citep{Eppinga2023, Mathias}. Analytical tractability of DSM and their ability to study asymptotic dynamics comes at the expense of models' ability to represent demographic heterogeneity and detailed spatial structure. 

ABM, on the other hand, is a computational tool where autonomous agents act and interact with each other by learning and adapting to changes and a simulated environment. ABM are designed to study the system- or macro-level states or dynamics that emerge from local, micro-level interactions between diverse human and non-human agents and the effect of emerging macro-level dynamics on micro-level interactions \citep{Schluter2021}. They offer enhanced flexibility in simulating how agents based on simple rules, perform certain actions and interact with each other and their environment. Through simulating the virtual world and experimenting with it, researchers can explore the unfolding of the system over time and study consequences of changes to the system, e.g. through a policy or an environmental change, or explore mechanisms that may of generated a phenomenon of interest \citep{Schluter2019a}. ABMs have been used to explore social-ecological traps in empirical case studies \citep{Brinkmann, Kim, Nhim} and offer the methodological advantage of focusing on the heterogeneity, human decision-making/agency, learning and interactions of agents within the social-ecological system \citep{Lippe, Schulze}. Due to their object-oriented nature ABM is generally more intuitive for nonmathematicians to use than DSM that are based on ordinary, partial and stochastic differential equations and mathematical techniques for their analysis \citep{An}. 

Reviews of modeling approaches reveal the strengths and weaknesses of DSM and ABM for exploring SES dynamics, particularly social-ecological feedbacks, features of complexity (such as spatial, temporal and organizational scales, nonlinearities and thresholds), human behavior or heterogeneity \citep{An, Filatova2013, Filatova2016, Schluter2019b}. Usefulness of models depends on their purpose, which can vary from understanding, theory building and exploring generic system behavior to predicting and providing management support \citep{Edmonds}. Models can serve as analytical, predictive, or exploratory tools, but they can also be a boundary object at the intersection of theory and empirics or in participatory processes. Model purpose influences the level of realism and details that need to be included, while its assumptions and conceptualization shape and constrain causal explanations that the model provides \citep{Banitz2022}. More details are not necessarily better, especially in complex systems with complicated causal feedback loops. Adding too much complexity too early in the modeling process can make an ABM difficult to understand and check for errors. Researchers might risk replacing poorly understood real life problems with hard to understand models \citep{Osullivan, Manzo}. Justifying that micro-level interactions and high levels of detail are really needed in ABM is considered a good modeling practice. Then again, reducing complexity when designing a DSM can hide important dynamics and lead to wrong conclusions. This practice is particularly dangerous in cases when spatial or demographic heterogeneity and interactions between individuals are important for the outcome, as in biological conservation or epidemiological modeling \citep{Wennergren, Grossmann, Anderies}, or when key interactions in the systems are not purely economic or ecological, but social-ecological in nature, as in poverty traps \citep{Lade2017} or in exploited ecosystems \citep{Fogarty}. 

Apart from the major challenges in representing complexity and heterogeneity and integrating processes at different levels or scales, such as micro-level human decision-making with meso- and macro-level environmental and economic processes, models need to explore whether dynamical patterns  found in social-ecological systems represent  traps, regime shifts or long transients \citep{Radosavljevic2023}. Ability of DSM to identify attractors and explore asymptotic dynamics and bifurcations is undisputed, but equilibrium thinking does not always provide adequate answers for SES research. Real-world systems are frequently affected by shocks and stochastic events and guided by processes with complex temporal and spatial dynamics that can prevent the system from having or reaching attractors. They can create complicated trajectories whose shape can be of importance for understanding behavior of the system. Exploring transient dynamics may be more relevant for understanding and managing ecosystems and SES \citep{Eppinga2021, Hastings, Morozov}. Unlike DSM, ABM are not designed to deal with asymptotic dynamics, but to compute a finite time series of the system states. ABM are limited in their ability to identify trap dynamics or regime shift or to explain how traps emerge even though time series may show trap-like patterns, but they are more effective in  exploring transient dynamics than DSM.

Strengths and weaknesses of ABM and DSM stated above make it impossible to choose the best model type to represent possible system configurations, answer multilayered research questions and fit various purposes at the same time. Combining modeling approaches to increase the spectrum of insights provided by the models has been proposed \citep{Filatova2013, Filatova2016, Schluter2019b}. Potential benefits of combined DSM-ABM and frameworks for creating hybrid models have been already explored \citep{Guerrero, Martin, Rahmandad, Swinerd}. According to \citet{Martin}, a hybrid approach of combining system dynamics and agent-based modeling can allow for the analysis of social-ecological dynamics at the aggregated level but also the role of human decision-making in influencing these dynamics. It also allows one to use the respective strength of each modeling type, i.e. the mathematical toolbox from DSM to explore stability and bifurcations and the ability to represent heterogeneity, space and agent behavior in the ABM. 

The aim of this paper is to explore how combining DSM and ABM can improve understanding SES dynamics and to provide a systematic procedure for leveraging the strengths of both approaches. \citet{Martin} coupled an ABM and system dynamics model and developed a hybrid model of freshwater lake to tackle some of these challenges. Through a stepwise procedure they systematically explored both the decoupled sub-models as well as the coupled full model analysis.  Our goal is similar to theirs, but our approach differs in that the DSM and ABM we create are not coupled but standalone models that build on each other. The insights generated in each model are used as inputs for model development or scenario generation for the respective other model. Iterative process of the ABM analysis, creating a DSM using inputs from the ABM, analyzing the DSM and switching back to the ABM using results of the DSM analysis creates a dialogue between the models. This dialogue, more than having a tight connection between the ABM and DSM, enables pushing the research frontier and expanding the set of questions that can be asked and insight that can be gained. We illustrate our method with the example of poverty traps and innovation in agricultural systems based on the Ag-Innovation ABM \citep{Sanga2023} and poverty trap models in agroecological systems \citep{Radosavljevic2020, Radosavljevic2021}. 

The starting point of our model combination is reducing complexity of the system representation through a systematic process that was based on the results of the ABM. This allowed us to create a DSM that describes the same system as the ABM, but with less details. Stability and bifurcation analysis of the DSM uncovered asymptotic dynamics of the system and identified traps and conditions under which they appear. However, the reduced complexity of the DSM, left questions related to heterogeneity of farmers open. Switching back to the ABM, while using insights from the DSM as guidelines, allowed us to reintroduce heterogeneity in the critical parts of the system and in turn get a better understanding of the role that farmers’ characteristics in combination with social-ecological dynamics play in the creation of poverty traps. Together, the DSM and ABM models allow for identification of solutions for prevention of poverty traps. The models perform autonomously but in synchrony, and leverage the strengths of both DSM and ABM approaches to provide insights into the system dynamics. 

The paper is organized as follows. In Section 2 we briefly introduce the case study that will be used as an example or illustration for the process of combining ABM and DSM. In Section 3, we specify a procedure for ABM-DSM co-development and analysis, explore what aspects of complexity and heterogeneity are changed when we move from one modeling paradigm to another, and investigate how this dialogue between the models provides new insights into asymptotic and transient dynamics of the system. The aim of the paper is not to present a complete overview of all relevant results that this model combination leads to, but to demonstrate the strength of both approaches when they are combined; an in-depth analysis of the combined modeling approach and their application to a case study will be published elsewhere. Main contributions of the DSM-ABM combination to understanding complexity, heterogeneity, transient and asymptotic dynamics and facilitating interdisciplinary research are presented in the discussion.

%%%%%%%%%%%%%%%%%%%%%%%%%%%%%%%%%%%%%%%%%%%%%%%%%%%%%%%%%%%%%%%%%%%%%%%
\section{Case study}
\label{Sec2}

Poverty traps are defined as situations where feedbacks between social and ecological systems create undesirable, unsustainable, or maladaptive states that are highly resistant to change \citep{Amparo, Barrett2003, Boonstra, Radosavljevic2021, Unruh}. In SES literature, poverty traps are closely related to the notion of bistability, i.e. the existence of two attractors, where one attractor represents a poor state and the other represents a well-being state. Works based on DSM study intertwined ecological, social, behavioral and biophysical processes that create traps \citep{Lade2017, Ngonghala, Radosavljevic2020, Radosavljevic2021}, but do not account for demographic and spatial heterogeneity. Poverty alleviation strategies are based on developing interventions that would move the system from poor to well-being basin of attraction. Lack of heterogeneity in the models allows only partial answers.

Recent work using ABM \citep{Sanga2023}, explored contributions of agricultural innovations to food security and escape from poverty in an agricultural system. Quantitative empirical data for the biophysical dimension and qualitative data and knowledge from a particular case in Mali, West Africa were used to develop an empirically-stylized model \citep{Sanga2020, Sanga2021} that features aspects of spatial and demographic heterogeneity. Simulations with the ABM revealed trap-like dynamics, however it was difficult to  provide definite answers whether the dynamics observed in the model were really creating traps and which levels interacted to create traps. These questions could be addressed by analyzing asymptotic dynamics of a DSM representing the same system, which gave inspiration for creating a model combination we present in this paper.
%%%%%%%%%%%%%%%%%%%%%%%%%%%%%%%%%%%%%%%%%%%%%%%%%%%%%%%%%%%%

\section{A procedure for combining DSM with ABM}

Our procedure for combining ABM and DSM has eight steps:  1) identifying causal explanations for the phenomenon of interest in empirical case studies and selecting theoretical and expert knowledge that provides insights into the studied system to inform the ABM, 2) ABM development and analysis, 3) identifying key assumptions for causal loop diagram (CLD) based on the results of the ABM, 4) selecting key variables and DSM formalization, 5) stability analysis of the DSM and explanations based on asymptotic dynamics, 6) bifurcation analysis of the DSM, structurally unstable systems and scenario development, 7) scenario based ABM simulation and analysis, and 8) assessment of leverage points and intervention design using the ABM and the DSM.  

\begin{figure}[h]
\centering
\includegraphics[width=\linewidth]{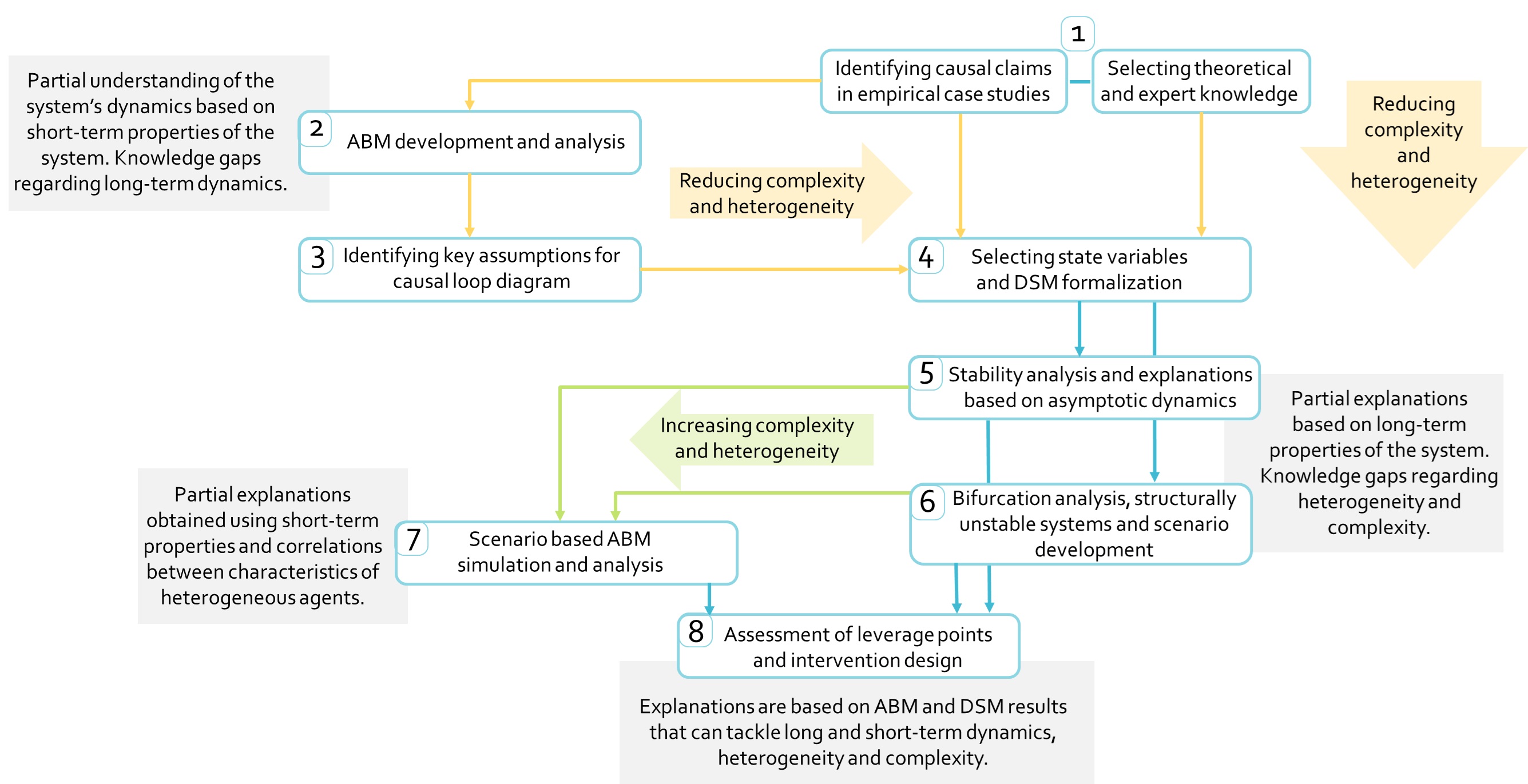}
\caption{\footnotesize  Illustration of the procedure for combining ABM and DSM. Yellow arrows represent parts of the modeling process that decrease complexity and heterogeneity of the studied system. Green arrows represent modeling processes that bring additional complexity and heterogeneity. Blue arrows stand for mathematical and analytical methods within DSM and ABM. Gray boxes contain explanations obtained from the models.}
\label{Figure1}
\end{figure}

We illustrate the procedure using poverty traps in an agricultural system with endogenously driven innovation presented in Section 2. Steps (1)-(8) are stated in the order in which we use them in this study (Figure \ref{Figure1}). This is in particular related to the pre-existing ABM that provides inputs and empirical basis for a DSM that we develop and analyze. The procedure can be applied to other case studies, systems and phenomena.  Its purpose is to guide model co-development, clarify contributions of each modeling approach to understanding dynamics of the chosen system, and facilitate dialogue between models. Order of steps in the procedure is less important than the insights they provide and it can be changed to fit the particular study or in cases when an ABM is not available and model co-development begins with a DSM.

%%%%%%%%%%%%%%%%%%%%%%%%%%%%%%%%%%%%%%%%%%%%%%%%%%%%%%%%%

\subsection{Identifying causal claims in empirical case studies and selecting theoretical and expert knowledge}
\label{Sec3.1}

The system conceptualizations for ABM and DSM include the tacit assumption that the entities and interactions considered in each model are relevant for representing the real world system and answering the research question. The model-based causal findings depend on context, which in the modeling sense means that a model inputs and the way the model is analyzed predefine its outputs \citep{Banitz2022}. Selecting appropriate empirical and theoretical inputs is common for both modeling techniques. What might differ in this process for ABM and DSM are the source of knowledge and model inputs (empirical observation/data are more often associated with ABM and theoretical knowledge and known functional forms with DSM) and the level of abstraction and aggregation, where we expect more abstraction and less heterogeneity and complexity for creating DSM.

%%%%%%%%%%%%%%%%%%%%%%%%%%%%%%%%%%%%%%%%%%%%%%%%%%%%%%%%%
\subsection{ABM description and analysis}
\label{Sec3.2}

The Ag-Innovation ABM \citep{Sanga2023} captures four key processes of agricultural innovation: development, dissemination, adoption, and diffusion of innovation involving producers and innovators. Producers own land, cultivate crops, form beliefs and desires about innovation, and adopt innovations. Innovators are agents that develop innovations based on pooled capital and knowledge sharing among producers. Producers interact with farmland by cultivating crops based on climate risk perception, which also influences their desires about innovations needed for their farmland.  Innovators interact with producers by capital pooling, assessment of innovation demand,  forming networks and disseminating information about developed innovations. The model also incorporates ecological interactions between crop and soil through soil fertility regulation. Full ODD protocol \citep{Grimm} for the model is in Appendix~A. 

The model was developed to explore how different mechanisms of innovation, particularly exogenous versus endogenous driven, affect outcomes related to food insecurity and income inequality. The analysis of the model shows that the endogenous mechanisms, contrary to expectations, lead to higher income inequality than exogenous mechanisms. This unexpected outcome may indicate the formation of a social-ecological poverty trap where producers pool capital for collectives to farmers develop production enhancing innovations that may harm their soils in the long run and further decrease crop production and income.  These dynamics closely resembled findings from a study by \citep{Radosavljevic2020}, which highlighted how interconnected social and ecological factors not only influence food production but can also contribute to the emergence of a poverty trap. The questions that the ABM could not answer was: How do these poverty traps emerge? Would understanding long term dynamics of such traps help us design efficient poverty alleviation strategies?

%%%%%%%%%%%%%%%%%%%%%%%%%%%%%%%%%%%%%%%%%%%%%%%%%%%%%%%%%%

\subsection{Identifying key assumptions for Causal Loop Diagram}
\label{Sec3.3}

The link between the ABM and the development of the DSM is established by constructing a causal loop diagram using the ABM assumptions and results. In the following steps, this CLD can be translated into an DSM, first by identifying key variables of the ABM and then listing how changes in one variable affects changes in other variables. This establishes the basis of causal understanding of processes within the system and guides DSM development.

\begin{figure}[h]
\centering
\includegraphics[width=0.5\linewidth]{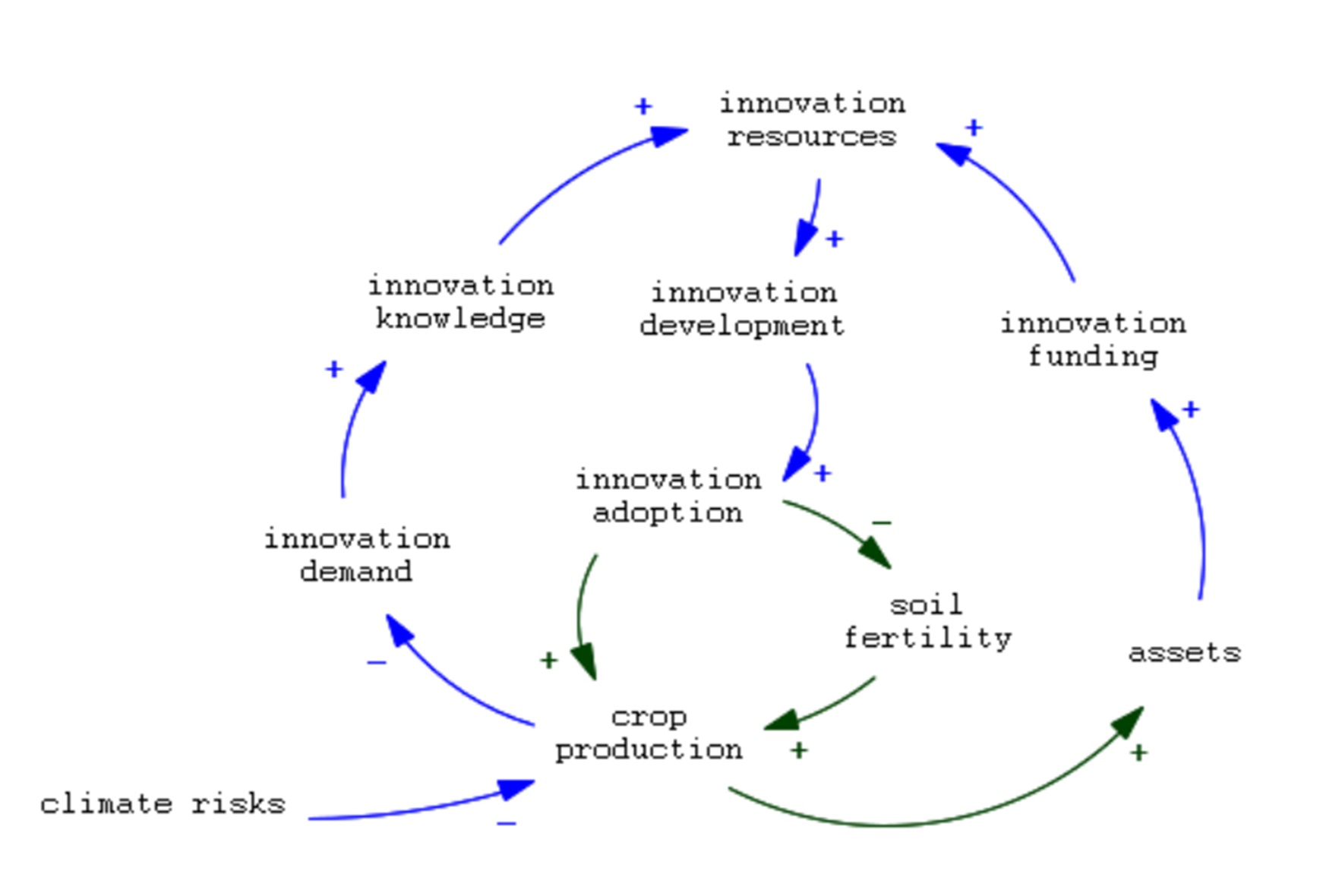}
\caption{\footnotesize Causal loop diagram based on structure of Ag-Innovation ABM. Arrows in blue represent variables operating at meso-level while arrows in green represent variables operating at micro-level.}
\label{Figure2}
\end{figure}

Based on the structure of the Ag-Innovation model (Appendix A), we identified eight key variables and parameters that allow for an understanding of the nature of relationships in the model: climate risks, crop production, innovation demand, innovation knowledge, innovation resources, innovation development, innovation funding, and soil fertility. The variables are related through following causal relations: 1) Increase in climate impact events leads to decrease in crop production, which in turn leads to an increase in innovation demand. 2) Increase in innovation demand increases innovation knowledge. 3) Decrease in crop production also leads to a decrease in assets. 4) Decrease in assets also leads to decrease in innovation funding. 5) Both innovation funding and innovation knowledge influence innovation resources. 6) Increase in innovation resources leads to increase in innovation development and innovation adoption. 7) Increase in innovation adoption leads to an increase in crop production and decline in soil fertility, which in turn leads to a decrease in crop production. Figure~\ref{Figure2} presents a causal loop diagram for the Ag-Innovation model. 

%%%%%%%%%%%%%%%%%%%%%%%%%%%%%%%%%%%%%%%%%%%%%%%%%%%%%%%%%%
\subsection{Selecting state variables and DSM formalization}
\label{Sec3.4}

Results from the previous step (Figure \ref{Figure2}) show the central role of crop production and highlight the importance of innovation for crop production. What this diagram does not include are ecological interactions between soil fertility and crop growth, and economic dynamics that represent farmers' use of financial assets for consumption and investment in agricultural production and innovation. To include missing ecological and economic dynamics, we complement results from the ABM with economic growth models \citep{Barro}, soil and nutrient dynamics \citep{Bunemann, DeAngelis, Drechsel} and poverty traps in agricultural context \citep{Barrett2003, Lade2017, Radosavljevic2020, Radosavljevic2021}.

\begin{figure}[h]
\centering
\includegraphics[width=0.5\linewidth]{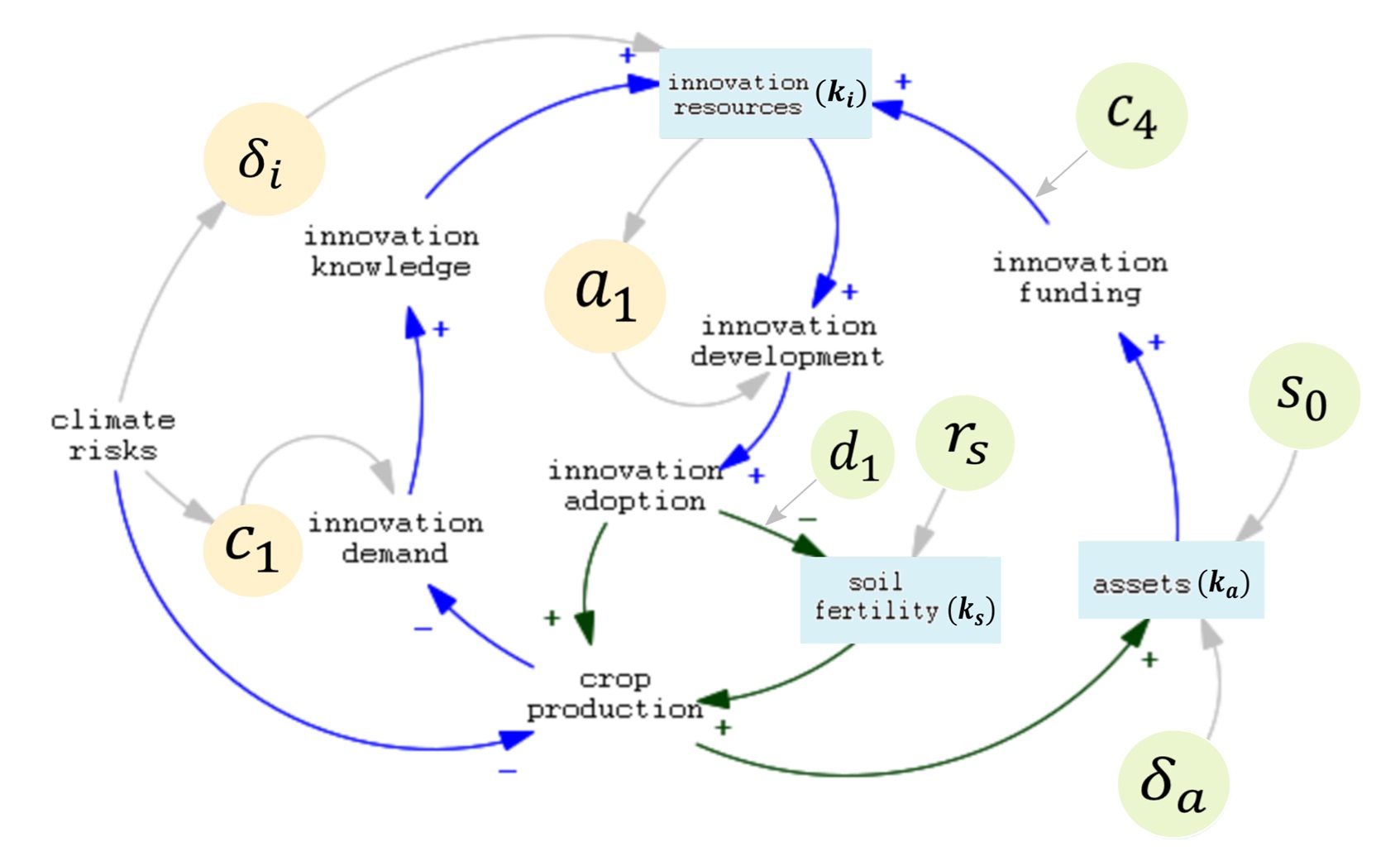}
\caption{\footnotesize Causal loop diagram of the DSM based on assumptions \ref{A1}--\ref{A8} and Figure \ref{Figure2}. State variables are given in blue boxes and parameters are in circles ($a_1$- innovation efficiency, $c_1$- innovation desire, $\delta_i$ - depreciation rate of innovation resources, $r_s$ - soil fertility recovery rate, $s_0$ - savings rate, and $\delta_a$ - assets’ depreciation rate). Yellow circles indicate DSM parameters coming from the ABM, while green circles indicate DSM parameters supported by ecological and economic literature. Full definition of parameters can be found in Table 1 in Appendix B. Blue arrows represent cross-level interactions. Green arrows represent individual-level interactions.}
\label{Figure3}
\end{figure}

Causal relations explored in the ABM and given in Figure \ref{Figure2} are translated into relations between state variables in the DSM. They are firstly summarized in the following assumptions:
\begin{enumerate}[label=($A_{\arabic*}$)]
    \item Innovation resources improve crop production by increasing productivity. \label{A1}
\item Innovation resources have a negative effect on soil fertility. \label{A2}
\item Decrease/increase in soil fertility decreases/increases crop production. \label{A3}
\item Higher productivity reduces innovation demand and funding and consequently its adoption. \label{A4}
\item Higher productivity increases farmers’ assets and consequently innovation funding. \label{A5}
\item Climate risk events decrease innovation resources (by reducing the effectiveness of technology) and decrease crop productivity. \label{A6}
\end{enumerate}
Following ecological and economic literature, we have additional assumptions for the DSM:
\begin{enumerate}[resume,label=($A_{\arabic*}$)]
    \item Future level of innovation resources depends on the current level of innovation resources. \label{A7}
    \item Soil fertility has logistic growth. \label{A8}
\end{enumerate}
The central role of crop production is preserved in the DSM, but since crop growth depends on soil quality and productivity depends on innovation and assets, we use assets, soil fertility and innovation resources as state variables in the DSM (Figure \ref{Figure3}, Table \ref{Table1}).

To define economic dynamics and represent farmers investment in endogenous innovations, we use assets $k_a$ as a state variable in the DSM. Assets are defined as farmers’ financial and physical capital and follow the neoclassical growth model and poverty traps models \citep{Barro, Radosavljevic2020, Radosavljevic2021}. The central part of agricultural activities is crop production. Figure \ref{Figure2} shows that crop production and innovation resources are connected by two different chains of mechanisms: 1) through innovation demand and innovation knowledge, and 2) through assets and innovation funding. The first chain describes non-financial mechanisms while the second describes financial mechanisms. Assumption \ref{A7} describes endogenous innovation, where future innovations are possible only if knowledge of the system is preserved within the community. If the community loses its knowledge, learning and innovation would have to start from the beginning. For these reasons, we use innovation resources $k_i$ as one of the state variables in the DSM (Table \ref{Table1}) and define it as financial and non-financial resources that are needed for innovation development. 

Interactions between the environment and agricultural activities are described using soil fertility $k_s$ as a state variable in the DSM. Its role in the system is defined by assumptions \ref{A2}, \ref{A3} and \ref{A8}, which means that without damaging effects of innovation, soil fertility follows logistic growth. We are aware that assumption \ref{A8} is a huge oversimplification of the concept of soil fertility, but the focus of the model is on exploring innovation, not soil dynamics. Soil fertility is sometimes defined as the amount of organic matter in the soil, which partially justifies why we use logistic function to describe its behavior.

Summarizing assumptions \ref{A1}--\ref{A8}, relations from Figure \ref{Figure3} and deliberations stated in this section, we arrive at formalization of the DSM as a system of ordinary differential equations:

\begin{equation}\label{model1}
    \begin{aligned}
    \frac{dk_a}{dt} &= \left(s_0+\frac{s_1k_a^m}{s_2+a_k^m}\right)\left(a_0+\frac{a_1k_i^n}{a_2+k_i^n}\right)k_a^{\alpha_a}k_s^{\alpha_s} - \delta_ak_a\\
    \frac{dk_s}{dt} &= r_sk_s(K-k_s)-\frac{d_1k_i^s}{d_2+k_i^s}k_ak_s \\
    \frac{dk_i}{dt} &= \left(c_1+\frac{c_2}{c_3+k_a}\right)\frac{c_4k_a}{c_5+k_a}\frac{k_i}{1+k_i}-\delta_ik_i.
    \end{aligned}
\end{equation}
Complete list of parameters and explanations about used functional forms of the system of differential equations are in Appendix B.

In order to ensure tractability, we chose to keep the number of state variables low in model (\ref{model1}). Three state variables are selected among eight variables of the ABM and causal relations between them are preserved (see blue fields in Table \ref{Table1}). Consequently, some mechanisms present in the ABM are changed in the DSM or several mechanisms are merged into one. For example, innovation demand and innovation knowledge that are used in the ABM are merged and represented by parameter c1 (innovation desire) in the DSM (see yellow fields in Table \ref{Table1}). Removing or simplifying mechanisms can lead to certain loss of complexity in the DSM, while assuming that all farmers have identical behavior, properties and spatial details results in loss of heterogeneity in the DSM. 

\begin{table}[ht]\footnotesize
\begin{tabularx}{\textwidth}{m{2cm}|m{7cm}|m{7cm}}
\toprule
\vspace{0.1cm} \bf{Variables} \vspace{0.1cm}  & \vspace{0.1cm} \bf{Definition in the ABM} \vspace{0.1cm} & \vspace{0.1cm} {\bf{Definition in the DSM}} \vspace{0.1cm}
\\\hline
\rowcolor{lightcyan}
Assets & Physical and financial capital endowment (food and income)  of producers &  \vspace{0.1cm} State variable $k_a$ represents farmers' financial and physical capital in model (\ref{model1}). Their dynamics is defined using parameters $s_0$ and $a$. 
\\\hline
\rowcolor{lightcyan}
Soil fertility &  Index representing soil quality and proportion of nutrients in plots owned by producers & \vspace{0.1cm} State variable $k_s$ denotes concentration of nutrients and organic matter in the soil in model (\ref{model1}). For details see Appendix B. \vspace{0.1cm}
\\\hline
\rowcolor{lightcyan}
Innovation resources &  Indicates the financial as well as non-financial resources (such as knowledge) needed for innovation development & \vspace{0.1cm} State variable $k_i$ in model (\ref{model1}) represents financial and non-financial resources needed for innovation development, e.g. innovation funding and knowledge. For details see Appendix B. \vspace{0.1cm}
\\\hline
\rowcolor{lemonchiffon}
\vspace{0.1cm} Crop \newline production \vspace{0.1cm} & \vspace{0.1cm} Total amount of crop produced by producers for sorghum, millet, maize and rice \vspace{0.1cm} & \vspace{0.1cm} Given as the crop production function in model (\ref{model1}). For details see Appendix B. \vspace{0.1cm}
\\\hline
\rowcolor{lemonchiffon}
Innovation demand & \vspace{0.1cm} Indicates the innovation desires and needs of producers, providing a direction for innovators for innovation development \vspace{0.1cm} & 
\\\cline{1-2}
\rowcolor{lemonchiffon}
\vspace{0.1cm} Innovation \newline knowledge \vspace{0.1cm} & \vspace{0.1cm} Indicates the knowledge of innovators of innovation desires and needs of producers \vspace{0.1cm} & \multirow{-2}{*}[3em]{\parbox{7cm}{Represented by parameter $c_1$ (innovation desire) that defines how assets through perception of current and desired outcomes of crop production affect innovation funding. See Figure \ref{Figure3}.}}
\\\hline
\rowcolor{lemonchiffon}
\vspace{0.1cm} Innovation \newline development \vspace{0.1cm} & \vspace{0.1cm} Types of innovation developed by innovators \vspace{0.1cm} &
\\\cline{1-2}
\rowcolor{lemonchiffon}
\vspace{0.1cm} Innovation \newline adoption \vspace{0.1cm} & \vspace{0.1cm} Types of innovation adopted by producers \vspace{0.1cm} & \multirow{-2}{*}[1em]{\parbox{7cm}{Represented by parameter $a_1$ denoting innovation efficiency. See Figure \ref{Figure3}.}}
\\\hline
\rowcolor{lemonchiffon}
\vspace{0.1cm} Innovation funding \vspace{0.1cm} & \vspace{0.1cm} Total pooled amount of allocated capital by producers \vspace{0.1cm} & \vspace{0.1cm} Parameter $c_4$ defines purely economic links between crop production and innovation resources. \vspace{0.1cm}
\\\hline
\rowcolor{lightgreen!30}
Environmental \newline damage & \vspace{0.1cm} Environmental damage potential with values “high” and “low” is used to design ABM scenarios in Section \ref{Sec3.7} \vspace{0.1cm} & Parameter $d_1$ defines maximal negative effects of innovation on soil fertility. See Figure \ref{Figure5}.
 \\\bottomrule
\end{tabularx}
\smallskip
\caption{\footnotesize Key variables and parameters for understanding the nature of relationships in the Ag-Innovation ABM and their comparison with key variables and parameters used in the DSM (model~\ref{model1}). Entries in yellow cells are variables in the ABM that are translated into parameters of the DSM. Entries in blue cells are state variables of the DSM. Entry in the green cell is a parameter of the DSM that is used for additional analysis and scenario development in the ABM.}
\label{Table1}
\end{table}

%%%%%%%%%%%%%%%%%%%%%%%%%%%%%%%%%%%%%%%%%%%%%%%%%%%%%%%%%%%%%%%%%%%%%
\subsection{Stability analysis and explanations based on asymptotic dynamics}
\label{Sec3.5}

Stability analysis identifies and classifies equilibrium points of a dynamical system and explores its asymptotic dynamics i.e. the dynamics that the system will enter  after a transient phase and maintain forever if not subjected to shocks or perturbations \citep{Radosavljevic2023}. A particularly important type of equilibria is attractor, defined as a state or set of states toward which all neighboring states converge. Systems with one attractor have only one long-term regime independent of the initial conditions. Systems with two or more attractor i.e. bistable (or multistable) systems, have two (or more) long-term regimes. These systems show path dependency because the future states of the system depend on the initial conditions.

\begin{figure}[h]
\centering
\includegraphics[width=0.3\linewidth]{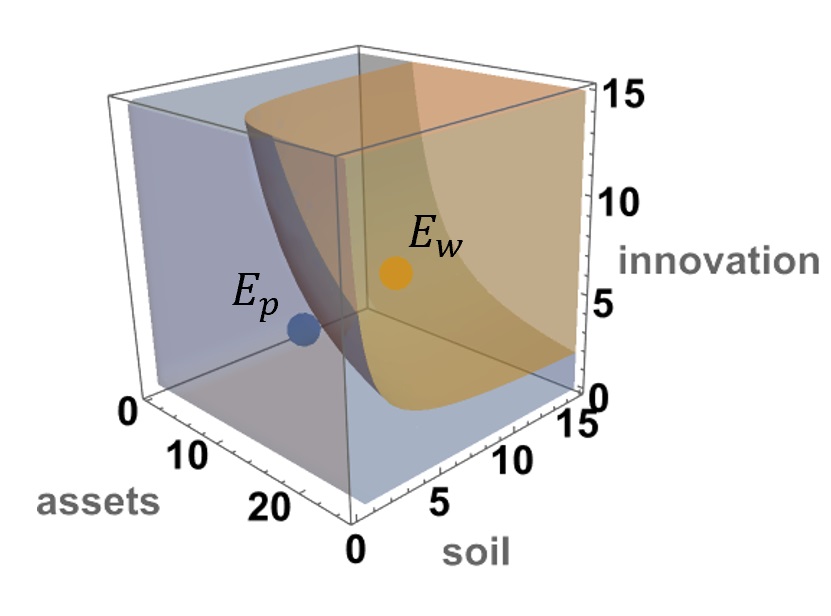}
\caption{\footnotesize Stability analysis for innovation with low environmental damage shows a bistable system. $E_p$ denotes the poor attractor, characterized by low assets and innovation levels and high soil fertility. Blue volume is the corresponding basin of attraction.  $E_w$ and yellow volume represent the well-being attractor and its basin of attraction. }
\label{Figure4}
\end{figure}

Stability analysis gives deeper insights into the kind of interventions that would enable reaching the desired attractor \citep{Radosavljevic2023}. Results of stability analysis of model (\ref{model1}) show that poverty alleviation is impossible in systems with a single attractor because all trajectories converge to the poor state and decline to permanent poverty is the only outcome regardless of initial conditions. Escape from poverty is possible in bistable systems if one of the alternative attractors defines a well-being regime. Visualizations, such as one in Figure \ref{Figure4}, point out what state variable values must be increased for trajectories to move to the well-being basin of attraction. 

Limitations of stability analysis, such as assuming constant parameters and not accounting for stochasticity or error in measurement can be amended by combining stability with bifurcation analysis, time series analysis and exploring properties of trajectories. 

%%%%%%%%%%%%%%%%%%%%%%%%%%%%%%%%%%%%%%%%%%%%%%%%%%%%%%%%%%%%%%%%%%%
\subsection{Bifurcation analysis and scenario development}
\label{Sec3.6}

Bifurcation analysis explores qualitative changes in systems dynamics due to changes in parameter values \citep{Kuznetsov}. It allows exploration of the whole parameter range and its results show how the number, character and location of equilibria change with parameters. For nonlinear models, bifurcation analysis is often done using software for numerical simulation (for example XPPAUT or Matcont).

\begin{figure}[h]
\centering
\includegraphics[width=\linewidth]{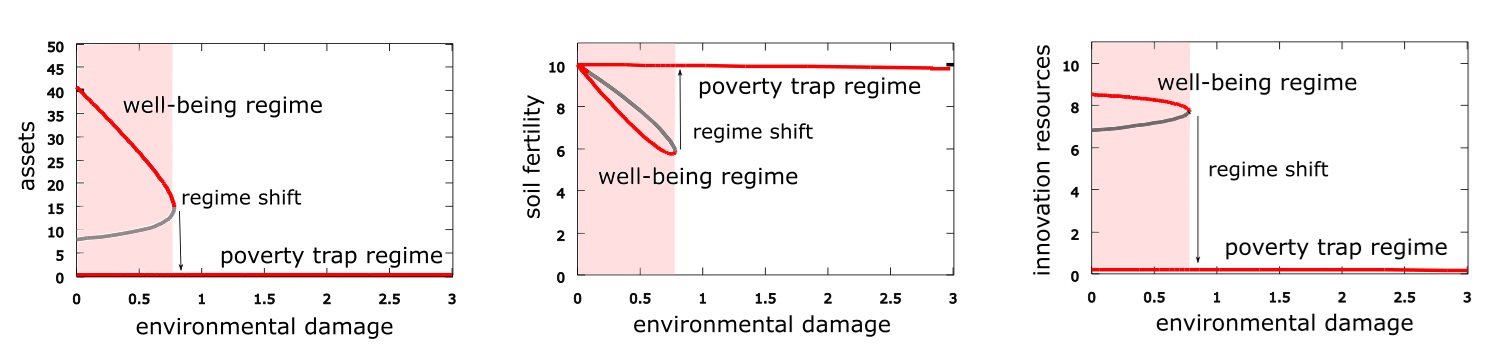}
\caption{\footnotesize Bifurcation analysis with environmental damage ($d_1$) as the bifurcation parameter shows bistability (light red area) for low values of parameter $d_1$ and monostability for high values of parameter $d_1$. In the area of bistability, well-being and poverty attractors coexist and escape from poverty is possible. A sudden jump (regime shift) occurs when parameter $d_1$ increases beyond its threshold value. After regime shift, the system is in the poverty trap regime and escaping from poverty is not possible.}
\label{Figure5}
\end{figure}

The DSM (\ref{model1}) in this paper has 17 parameters (Table 1 in Appendix B) and bifurcation analysis can be done for each of them or for their combinations. Development literature points out that increasing productivity through intensification often comes at the expense of the environment \citep{Lade2017}, which is why the extent of environmental degradation needs to be considered in designing interventions. Assumption \ref{A2} in Section \ref{Sec3.4} stated that innovation has a negative effect on soil fertility. First step in our analysis is therefore to explore the consequences of innovation induced environmental damage, which is defined by parameter $d_1$ in system (\ref{model1}). Bifurcation diagrams in Figure \ref{Figure5} show qualitative changes in the system’s dynamics for various values of parameter $d_1$ i.e. for innovations that produce various levels of environmental damage.

Attractors with high values of assets and innovation resources, but low values of soil fertility represent the well-being regime (Figure \ref{Figure5}). Poverty trap regime is characterized by attractors with almost zero assets and innovation resources, but high soil fertility. Figure \ref{Figure5} shows two outcomes depending on the value of environmental damage: a bistable regime in which well-being and poverty trap regimes co-exist (light red area) and a monostable regime where only poverty trap exists. These insights led to creation of two scenarios: 1) Gentle innovation that causes low environmental damage ($d_1<0.8$), and 1) Strong innovation that causes high environmental damage ($d_1\geq 0.8$). 

\begin{figure}[h]
\centering
\includegraphics[width=0.3\linewidth]{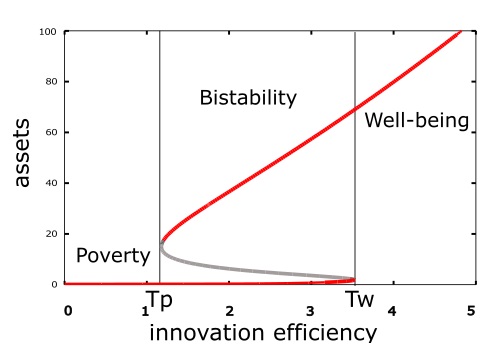}
\caption{\footnotesize Bifurcation analysis with respect to innovation efficiency (parameter $a_1$) assuming innovation with low environmental damage (low value of parameter $d_1$). $T_p$ and $T_w$ denote tipping points for poverty and well-being respectively. Poverty is the only outcome if $a_1<T_p$. Escape from poverty is enable by bistability of the system and it is possible if  $T_p<a_1<T_w$. Well-being is the only outcome in $a_1>T_w$.}
\label{Figure6}
\end{figure}

To gain a full understanding of the dynamics of innovations and their effects on the agricultural system, bifurcation analyses for both scenarios using parameters $a_1$ (innovation efficiency), $c_1$ (innovation desire) and $c_4$ (innovation funding) are needed, but it is beyond the scope of this paper to present all details of this investigation. We present only bifurcation diagram for innovation efficiency (parameter $a_1$) in Figure \ref{Figure6} due to its importance for the following sections of the paper.

Figure \ref{Figure6} indicates that innovation efficiency plays an important role for escape from poverty because the well-being attractor exists only for intermediate or high levels of parameter $a_1$. Poverty is the only outcome for low values of parameter $a_1$.

Both stability and bifurcation analysis explain behavior of DSM that is based on simplified assumptions (e.g. reduced number of processes in the system, no demographic and spatial heterogeneity). These analyses can explain how cross-level interactions create traps and help design effective poverty alleviation measures, but for a system based on average values of parameters for all farmers. The following question remains: What are characteristics of farmers that risk getting into poverty traps? To answer it, heterogeneity in farmers' characteristics needs to be reintroduced and explored in the ABM. 

%%%%%%%%%%%%%%%%%%%%%%%%%%%%%%%%%%%%%%%%%%%%%%%%%%%%%%%%%%%%%%%%%%%%
\subsection{Scenario-based ABM simulation and analysis}
\label{Sec3.7}

Bifurcation analysis with respect to the environmental  damage (Figure \ref{Figure5}) uncovered two distinct regimes (bistable and monostable), while additional bifurcation analysis with respect to innovation efficiency indicated that the well-being attractor exists only for medium or high values of parameter $a_1$ (Figure \ref{Figure6}). In both cases bifurcation parameters represent cross-level interactions between meso-level, where the innovation is created, and micro-level, where it is applied (for definition of parameters see Table 1 in Appendix B). Moreover, stability analysis of a bistable DSM (Figure \ref{Figure4}) gave a broader overview of the direction of external interventions that would enable the system to escape from poverty traps. The results explained the emergence of traps due to endogenous mechanisms and showed what levels interacted to create traps under assumption that all farmers are identical, but the lack of heterogeneity in farmers characteristics in the DSM prevented getting insights into who were more likely to be in the poverty states and which of their characteristics led them to that state. 

In order to assess implications of farmers' diversity, we switched back to the ABM under the guidance of the DSM. The ABM makes a distinction between two agents: producers and innovators and describes each of them with a set of characteristics (or attributes). In this paper we focus only on innovator agents who are described using four attributes: innovation capital, capital efficiency, knowledge efficiency, and innovation demand (Table 1 and Appendix A).

\begin{figure}[h]
\centering
\includegraphics[width=\linewidth]{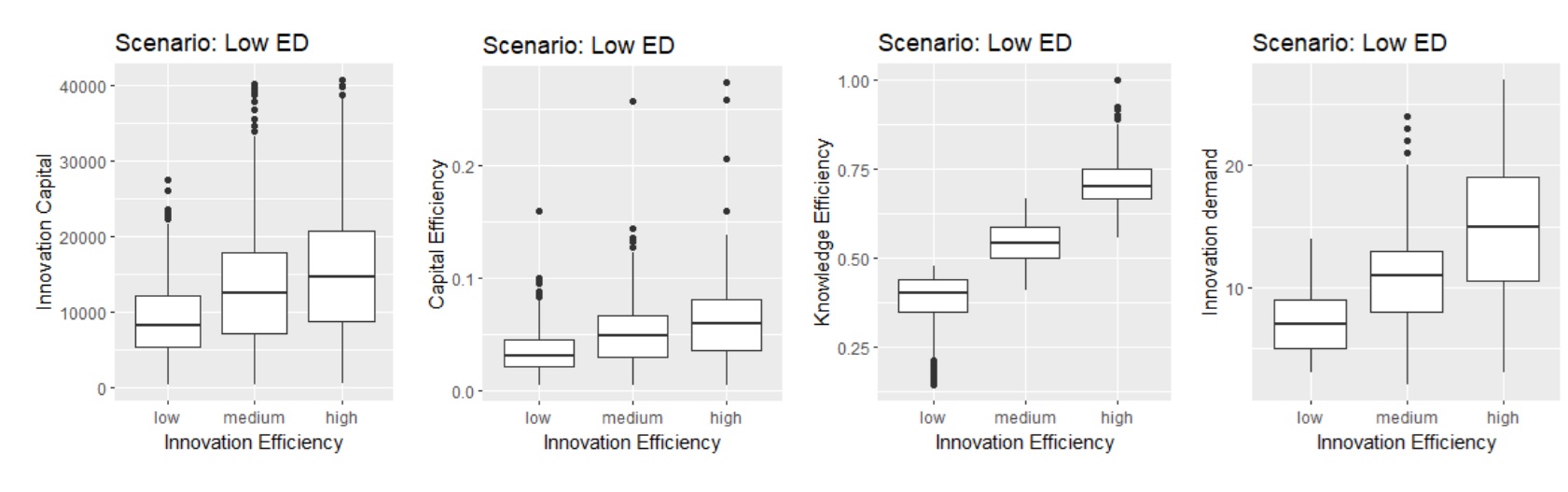}
\caption{\footnotesize From left to right are box plots representing distribution of innovation capital, capital efficiency, knowledge efficiency and innovation demand for different levels in innovation efficiency (parameter a1 in the DSM) for innovation with. low environmental damage (Low ED).}
\label{Figure7}
\end{figure}

Scenarios can be defined as a set of critical experimental parameters or “if-then” propositions that explore the consequences of a range of driving force assumptions on model outcomes \citep{Alcamo}. To account for environmental damage (Figure \ref{Figure4}), we ran two scenario analyses in the ABM with an index ‘environmental damage potential’ set at ‘high’ and ‘low’ respectively. The Ag-Innovation model was simulated for 200 time steps under the two scenarios and we captured data on changing attributes of individual agents at each time step.  We analyzed the scenario analysis data to identify characteristics of innovators who were likely to have lower innovation efficiency levels and hence, experience poverty traps as identified in the bifurcation analysis. Box plots in Figure \ref{Figure7} were used for graphical representations of the distribution and central tendency of the innovator agents attributes for low, medium, and high levels of innovation efficiency. 

We found that under innovations with low environmental damage, innovators with higher levels of innovation resources, capital efficiency, knowledge efficiency and innovation demand are likely to have higher innovation efficiency. Escape from poverty traps is possible at low and medium levels of innovation efficiency when there is an increase in all four factors; innovation resources, capital efficiency, knowledge efficiency, and innovation demand. While these results are not surprising, they exemplify how the ABM can be used to explore the implications of the results of the DSM when accounting for agent heterogeneity. Technical and abstract term “innovation resources” used in the DSM is translated into four characteristics of farmers that are much closer to the empirical case study and are easier to grasp from a practitioners and stakeholders perspective.  A full presentation of the ABM results, including the dynamics of producer and innovator agents for different bifurcation parameters from the DSM (such as innovation desire or funding) is beyond the scope of this paper and will be published elsewhere.

%%%%%%%%%%%%%%%%%%%%%%%%%%%%%%%%%%%%%%%%%%%%%%%%%%%%%%%%%%%%%%%%%%%%%
\subsection{Assessment of leverage points and intervention design}
\label{Sec3.8}

\begin{figure}[h]
\centering
\includegraphics[width=\linewidth]{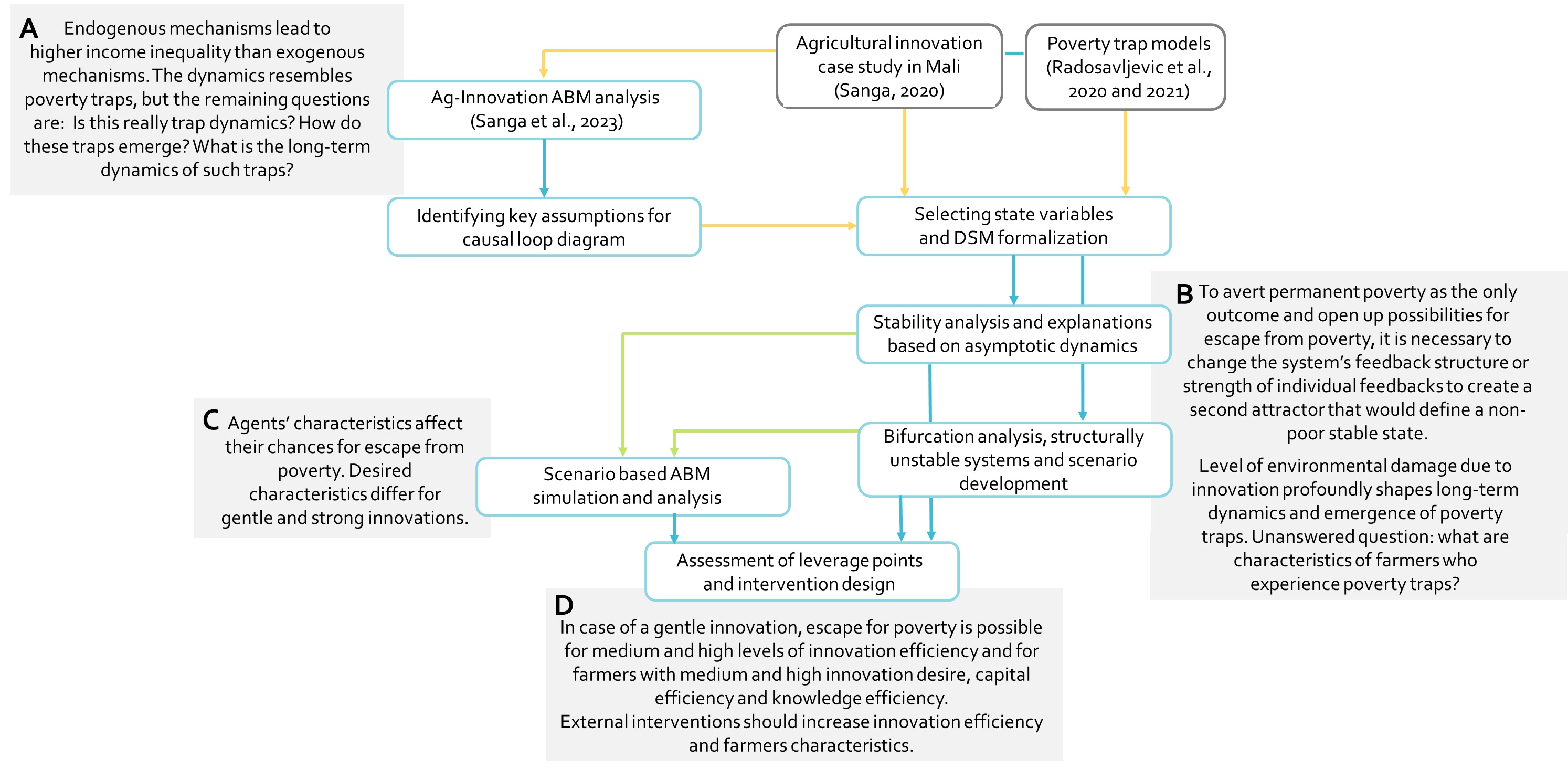}
\caption{\footnotesize  Combination of DSM and ABM applied to poverty traps study}
\label{Figure8}
\end{figure}

The final step in the process of combining models is analysis and synthesis of insights from the DSM and ABM obtained in previous steps. Visual summary of the process in Section \ref{Sec3.1} to \ref{Sec3.8} is in Figure \ref{Figure8}.

Bifurcation analysis (Figure \ref{Figure5}) shows that to prevent permanent poverty as the only outcome and open up possibilities for escape from poverty, it is necessary to change the system’s feedback structure or the strength of individual feedbacks. The aim of the system reconfiguration is to create a second attractor that would define a well-being stable state. Assessment of leverage points therefore relies on understanding what kind of action is needed to: 1) move initial points from poor to well-being basin of attraction, 2) to change the shape or size of the basin of attraction so that initial conditions converge to the well-being attractor, or 3) to create new well-being attractor if it does not exist. Lade et al. (2017) distinguish three types of interventions focusing on pushing the initial conditions to desired basin of attraction (type 1), changing strength of feedbacks in the system (type 2) and reconfiguring the system and creating new attractors (type 3). The shape of basins of attraction in Figure 4 can help evaluate usefulness of external interventions of type 1 and resilience of the well-being attractor under shocks. It can also help design intervention of type 2 that aims to increase the size of the well-being basin of attraction so that poor initial conditions become well-being initial conditions.  Finally, in case when the system has only a poor attractor, bifurcation analysis can show if changing parameter values would create a non-poor attractor. This could help design interventions of type 3 that involve system reconfiguration.

Analyses of the DSM (see Box B in Figure \ref{Figure8}) are conducted under the assumption that all farmers have the same characteristics. In the context of this study, heterogeneity means differentiating between producers and innovators and then considering characteristics of individual producers (e.g. capital, soil fertility, crop production)  or innovators (e.g. capital efficiency, knowledge efficiency or innovation demand). The DSM did not include these characteristics because functional forms that would describe them were unknown due to the lack of data about social-ecological interactions. In order to answer the question about characteristics of people who are likely to get into poverty, we shifted to the ABM.

The scenario analysis from the ABM allowed us to take a closer look at characteristics of agents and identify the specific type of interventions that would influence them (Box C in Figure \ref{Figure8}). Figure~\ref{Figure7} shows positive correlation between innovation efficiency and innovation resources, capital efficiency, knowledge efficiency and innovation demand. By highlighting the emergent behavior of interacting agents, the ABM can help us identify interventions that improve the initial conditions of producers and innovators that lead to poverty traps in the long run. These interventions act at a more granular level by focusing on how to increase capital efficiency, knowledge efficiency or innovation demand.  

Summarizing results of DSM and ABM analyses in the previous steps and contextualizing insights, we are able to provide answers to the research question that surpass answers obtained by analyzing single models (see  Box D in Figure \ref{Figure8}).

%%%%%%%%%%%%%%%%%%%%%%%%%%%%%%%%%%%%%%%%%%%%%%%%%%%%%%%%%%%%%%%%%%%%%%%%
\section{Discussion}

This paper develops and presents a systematic procedure for combining DSM with ABM in an iterative manner with the aim to generate deeper and more nuanced understanding of complex SES dynamics (Figure \ref{Figure1}). The procedure developed here is illustrated by the poverty trap case study (Section \ref{Sec2}), but can be applied to other ABM-DSM combinations aiming to explore different systems and phenomena. 

Differences between DSM and ABM in their conceptualization, level of details, methods for model analysis and how they treat complexity, heterogeneity, asymptotic or transient dynamics sometimes leave the impression that the modeling approaches are in opposition and that researchers need to choose one. In contrast to this, we think that the approaches are more complementing than opposing, the differences between them are more subtle and context dependent in comparison to how they are typically presented in the literature, and that usefulness of models is determined by the research questions, model purpose, available data, and even preferences of a modeler. We use analysis in Section 3 (Figure \ref{Figure8}) as an example, but our reflections on key contributions and challenges of the DSM-ABM combination are applicable beyond our case study.

\subsection{Moving from complex to simple and back}

Ability of a modeling approach to represent complexity is usually assessed along several dimensions, such as its capacity to represent and analyze multi-level systems, human behavior, nonlinearity, heterogeneity and thresholds. Both DSM and ABM are to varying degrees successful in addressing these dimensions, but results in Section \ref{Sec3.3} and \ref{Sec3.4} highlight another dimension of complexity that is of importance for DSM analysis: the number of elements in the CLD and properties of feedbacks and feedback loops between them. It is well-known that feedback loop character (positive or negative) gives necessary, but not sufficient conditions for multistability and periodicity. However, additional properties of single feedbacks within a feedback loop are often needed to ensure existence of multiple attractors or oscillations in the system. These might include nonlinear or nonconvex functional forms (e.g. Holling type II and III functional responses respectively) or functional forms that depend on several variables and cannot be written as products of simpler terms (e.g. Beddington-DeAngelis functional response) \citep{Radosavljevic2023}. A complete DSM formalization can therefore require additional information about functional forms and parameters that were not included in the ABM. 

Comparing CLD in Figure \ref{Figure2} and \ref{Figure3}, for example, we see that reducing complexity of the ABM involved merging several feedbacks into one and converting some of the ABM variables into parameters in the DSM (parameters in yellow circles in Figure \ref{Figure3} and yellow cells in Table \ref{Table1}), but also introducing new parameters in order to define interactions between variables in the DSM (parameters in green circles in Figure \ref{Figure3} and green cells in Table \ref{Table1}). The reasons for merging several feedbacks into one in DSM are the lack of quantitative and qualitative data and theoretical knowledge about social-ecological processes needed for defining proper functional forms for the DSM. There is also the lack of mathematical models for social-ecological interactions, or the uncertainty if a sequence of mechanisms produces sufficiently different dynamics than the single mechanism to justify additional data collection and more complicated model analysis.

Often the main reason for reducing complexity is the need to keep DSM tractable by keeping a low number of state variables because it allows better understanding of the model and facilitates experimenting with its structure. Unsurprisingly, this simplification may lead to misrepresentation of the system and ignoring potentially important interactions, which is why it needs to be treated with care, e.g., by ensuring better connection between theory and empirics, providing relevant social-ecological data or combining research methods. The iteration between ABM and the DSM helps ensure that the simplified system representation in the DSM includes the most important processes and facilitates an understanding of system dynamics that is not possible with more complex models. In addition, the simpler model enables easy and extensive experimenting with different system structures, which can only be done in a limited way with an ABM.

Further, ABM are typically designed to represent interactions among individual agents and agents’ properties, but do not necessarily involve all details of processes that are needed for the DSM. The reasons for this vary from the choice of research question, focus and conceptualization of the ABM, methodological needs and properties of ABM, but also to disciplinary traditions from which the model steams. Mathematical modeling is common in ecology or economics and there are well developed mathematical building blocks. In other disciplines, such as sociology, psychology or human geography, mathematical modeling is less prominent. While relevant knowledge exists, it is rarely formalized or organized in such a way to fit DSM \citep{Anderies}.

In our case study, the ABM was focused on human behavior and sufficient levels of details about soil and assets dynamics were missing.  Additional parameters in the DSM (in green circles in Figure~\ref{Figure3}) were incorporated from known ecological and economic literature. The DSM therefore inherits a social-ecological component from the ABM, adds ecological and economic components that surpass the ABM, and directs further ABM development and analysis (Section \ref{Sec3.7}). The ABM, on the other hand, guides the inclusion of key innovation dynamics and processes in DSM conceptualization and incorporates  details on individual and collective behavior of agents with heterogeneous characteristics that were not included in the DSM. The iteration between the two types of model thus allowed to broaden the scope of social and ecological dynamics that were considered when studying a SES, going beyond those that are commonly the focus of a particular modeling approach. Combining models not only creates a more nuanced understanding of the studied system, but inspires further model development, facilitates interdisciplinary work and brings closer different disciplinary traditions. 

\subsection{Levels of aggregation and heterogeneity} 
\label{Sec4.2}

In general case, different conceptualizations and sources of knowledge used for creating models can lead to creating ABM and DSM with different focus, levels of details and number of variables. It is often stated that DSM represents a system using aggregate state variables and feedbacks between them, while ABM represents a system through agents, environments and interactions at a disaggregated, micro-level. In different disciplines, the term \lq aggregation\rq\   has slightly different meaning, but usually involves two levels of organization, slow and fast time scales, and mathematical method to obtain aggregate variables. For example, in ecology, aggregation is used to reduce the number of variables describing individuals (micro variables) to obtain a few global variables representing groups of individuals. Global variables evolve on a higher organizational level and on a slow time scale \citep{Auger2000}. Aggregation methods in DSM start from high-dimensional systems and use the center manifold theorem to create systems of low dimension and verify that aggregated variables correspond to disaggregated variables \citep{Auger2008}. In economics, aggregation is about modeling the relationship between individual (micro) behavior and aggregate (macro) statistics, where macro variables are often seen as sums of micro variables.  

When it comes to ABM-DSM combination, the situation can be different because ABM are not always defined as systems of differential equations and aggregation principles typical for DSM or ecological modeling based on differential equations are not applicable. Instead, aggregation can be seen as a change in conceptualization and reduction of granularity of state variables that preserves time scale and organizational level as well as the average values of agent characteristics. For example, comparison between Figure \ref{Figure2} and \ref{Figure3} and Table \ref{Table1} show that both models represent variables at micro and meso levels, but that differences in granularity come from the lack of heterogeneity within the variables in the DSM. In other words, state variables of the DSM do not stand for aggregated, but averaged values. The DSM assumes that all individuals are identical, the space is homogeneous and interactions happen simultaneously without any specific spatial or temporal pattern. The ABM, on the other hand, explores macro-level patterns in income inequality and food security based on interaction of macro and meso level agents with heterogeneous demographic, economic, behavioral and  ecological  characteristics. 

Switching between ABM and DSM allows exploring consequences of variation in their conceptualizations and heterogeneity. The stability and bifurcation diagrams from DSM analyses helped identify which empirical details might matter and which ones not, thus contextualized heterogeneity and pointed to which characteristics of the agents should be more closely investigated in the ABM. For example, results of stability analysis in Figure \ref{Figure4} indicated that the well-being attractor is more vulnerable to decreasing innovation resources and assets than soil fertility, which highlighted the importance of agents’ characteristics for poverty alleviation. Bifurcation analysis in Figure \ref{Figure6} shows that bistability exists at medium levels of innovation efficiency, higher innovation efficiency leads to well-being, while low innovation efficiency leads to poverty. Relying on agent heterogeneity, the ABM allows identification of critical agents’ characteristics associated with low and medium levels of innovation efficiency and assessing which agents are more likely to end up in poverty in the long term. Together, the DSM-ABM generates insights into the design of real-world interventions that would potentially allow for escape from long-term poverty traps. Since the ABM is closely informed by empirics, these insights are more grounded  and applied to reality and validate the theoretical assumptions in the DSM. 

\subsection{Attractors, asymptotic and transient dynamics} 

Rich structure of a model in combination with simulations and time series analysis make ABM excellent for exploring short term dynamics of SES. The same richness, on the other hand, prevents investigating models’ asymptotic behavior and determining robustness of results. Questions that ABM is limited to answer are those related to existence and properties of equilibrium states and long-term dynamics of the system, such as regime shifts, emergence of social-ecological traps and possibilities for their escape. The benefit of reduced heterogeneity and complexity of DSM is that it facilitates manipulation of system causal structure and enables discovering key interactions, parameters and thresholds \citep{Schluter2024}, including those that create regime shifts and traps. 

In our case study, the ABM showed trap-like patterns suggesting that farmers might be caught in a poverty trap, but it was unable to verify the robustness of the result and left questions unanswered about emergence and creation of the cross-level poverty traps.  Stability analysis in the DSM showed existence of attractors and visualized basins for attraction (Figure \ref{Figure4}), which gave insights into undesired resilience of the poverty trap and possible leverage points. Additional clarity came from bifurcation analysis that determined parameter threshold values, described system dynamics for wide intervals of parameters, and helped dealing with uncertainty and errors in measurement. 

In continuous autonomous dynamical systems and in cases when parameters do not lead to bifurcations or when their values are far from thresholds, small changes in parameter value will not create qualitatively different dynamics. It is therefore safe to keep paremter values constant and shift focus to other elements of the system.  This reasoning, supported by bifurcation analysis in Figure \ref{Figure5}, led us to creating two scenarios in the ABM representing innovatons with low or high environmental damage. Within each scenario, the ABM was used to explore consequences of farmers' heterogeneity for their vulnerability to poverty. Using simulations and time series, ABM further clarified and contextualized specific interventions that would prevent poverty traps from occurring in the system (Section \ref{Sec3.7}). Thus, the ABM-DSM combination gave answers to questions that either of the models alone could not.

\subsection{Facilitating interdisciplinary work}

Creating ABM-DSM combination is not a linear process although we presented the procedure in eight steps (Section \ref{Sec3.1} to Section \ref{Sec3.8}). Creating a model combination capable of answering multi-layered research questions is far from trivial and it requires expert knowledge on ABM, DSM and on the empirical basis of the case study. In the process that illustrates our reasoning here, we were a dynamical system and two agent-based modelers who worked closely together to facilitate this iterative procedure of going back and forth. Each participant brought in their respective knowledge on the methods, but also on the empirical case study that informed the ABM. 

Having a dialogue between modelers was necessary for creating a dialogue between models. Each member of the team had to learn about both methods to be able to engage in the process of model combination. Challenges involved stepping out of a comfort zone of familiar model conceptualization and analysis and shifting methodological focus. For DSM, focus was on switching between asymptotic and transient behavior in order to learn about system dynamics. From an ABM perspective, it was more about learning about emergent properties using agent heterogeneity.

Apart from answering research questions, benefits of the process included gaining a shared understanding and better knowledge of the methods and case study. In the beginning of the process, we used literature on ABM-DSM comparison, where  comments regarding DSM were based on the theory of smooth, autonomous, continuous-time dynamical systems \citep{Kuznetsov, Wiggins}. The DSM has been judged without considering discontinuous, delayed and stochastic dynamical systems \citep{Bernardo, Jeffrey, Smith, Arnold} or dynamical systems game theory \citep{Akiyama2000, Akiyama2002}. It has also been stated that bifurcation analysis cannot be done for ABM, but \citep{Thomas} develop a method for conducting bifurcation analysis for equation-free models which can improve parameter estimation and understanding of ABM. Similarly, \citep{Bittermann} are exploring possibilities for constructing a stability landscape with ABM. Discussion in Section \ref{Sec4.2} points out that the term \lq aggregation\rq\ can have variations in meaning and should be used carefully in ABM-DSM context since it is not always clear what it entails. Thus, generalized statements about strengths and limitations of either modeling approach often cannot be justified and such claims can reflect knowledge gaps or indicate research frontiers more than actual properties of the approaches.

There is no blueprint showing how to make useful model combinations because the usefulness of models depends on the research question, model purpose, available data and preferences and skills of the researcher. DSM and ABM approaches are not in opposition and the differences between them are context dependent. Combining methods through an iterative and collaborative process can, however, be very rewarding, both in terms of enhanced understanding of complex system dynamics and in terms of developing more powerful tools to explore and explain social-ecological dynamics. 

%%%%%%%%%%%%%%%%%%%%%%%%%%%%%%%%%%%%%%%%%%%%%%%%%%%%%%%

\section*{Author credit statement}
Sonja Radosavljevic: Conceptualization, Methodology, Formal analysis, Visualization, Writing - original draft, Writing - review \& editing, Funding acquisition, Project administration. Udita Sanga: Conceptualization, Data curation, Methodology, Software, Visualization, Funding acquisition, Writing - review \& editing. Maja Schlüter: Conceptualization, Methodology, Writing - review \& editing, Funding acquisition.

\section*{Software and data availability}
Name of the software: AG-Innovation agent-based model

Developer: Udita Sanga, udita.sanga@tufts.edu

Program language: NetLogo

Software availability: 
\href{https://www.comses.net/codebases/80397098-9368-40ab-bb01-56b5f929ea04/releases/1.0.0/}{Model download}

Appendix A contains ODD for AG-Innovation ABM.
\newline
\newline
Appendix B contains mathematical details of the DSM and .ode file that was used for bifurcation analysis. No additional data was used for DSM analysis.

\section*{Acknowledgements} 
Funding from the Swedish Research Council FORMAS (grant number 2021-0184) is gratefully acknowledged. SR and MS acknowledge support from the Swedish Research Council (grant No 2018–06139). We are also thankful to Volker Grimm for discussions and comments that helped us improve the paper.

%%%%%%%%%%%%%%%%%%%%%%%%%%%%%%%%%%%%%%%%%%%%%%%%%%%%%%%%%%%%%%%%%%%%%%%%%%%%%%%%%

\bibliographystyle{plainnat}
\bibliography{bibtex}

\iffalse

\fi

\end{document}